\documentstyle[sprocl]{article}

\input{epsf}

\def\Journal#1#2#3#4{{#1} {\bf #2}, #3 (#4)}

\def\be{\begin{equation}}
\def\ee{\end{equation}}
\def\bea{\begin{eqnarray}}
\def\eea{\end{eqnarray}}
\begin{document}

\title{THE SPHERICAL COLLAPSE MODEL IN A UNIVERSE\\
WITH COSMOLOGICAL CONSTANT}

\author{E. L. \L OKAS}

\address{Copernicus Astronomical Center, Bartycka 18, 00--716 Warsaw,
Poland\\E-mail: lokas@camk.edu.pl}

\author{Y. HOFFMAN}

\address{Racah Institute of Physics, Hebrew University, Jerusalem 91904,
Israel\\E-mail: hoffman@vms.huji.ac.il}

\maketitle\abstracts{We generalize the spherical collapse model for the
formation of dark matter halos to apply in a universe with arbitrary
positive cosmological constant. We calculate the critical condition for
collapse of an overdense region and give exact values of the characteristic
densities and redshifts of its evolution.}

\section{Introduction}

The spherical collapse model was first developed by Gunn \& Gott~\cite{gg}
for a flat universe with no cosmological constant. It assumes that the
process of formation of bound objects in the universe can be at first
approximation described by evolution of an uniformly overdense spherical
region in otherwise smooth background. Despite its simplicity, the model
has been widely used to explain properties of a single bound object
via extensions such as the spherical infall model as well as statistical
properties of different classes of objects via Press-Schechter-like
formalisms. Given the presently mounting evidence for cosmological
constant we generalize the model here to include its effect.

\section{The cosmological model}

The evolution of the scale factor $a=R/R_0=1/(1+z)$ is governed by
equation
\begin{equation}       \label{th1}
    \frac{{\rm d} a}{{\rm d} t} = \frac{H_0}{f(a)} , \ \ \
    {\rm where} \ \ \
    f(a) = \left[ 1+ \Omega_0 \left(\frac{1}{a}
    -1\right) + \lambda_0 (a^2 - 1) \right]^{-1/2}.
\end{equation}
The parameters $\Omega_0$ and $\lambda_0$ are the present values of
$\Omega$ and $\lambda=\Lambda/(3 H^2)$ where $\Lambda={\rm const}$ is
the standard cosmological constant. The evolution of these parameters with
redshift $z$ is given by $\Omega(z) = \Omega_0 (1+z)^3 [H_0/H(z)]^2$
and $\lambda(z) = \lambda_0 [H_0/H(z)]^2$ where
$[H(z)/H_0]^2 = \Omega_0 (1+z)^3 - (\Omega_0 + \lambda_0 -1)(1+z)^2 +
\lambda_0$.

The linear growth of density fluctuations is given by
\begin{equation}    \label{th6}
    D(a) = \frac{5 \Omega_0}{2 a f(a)} \int_{0}^{a} f^3 (a) {\rm d} a
\end{equation}
where $f(a)$ was defined in Eq.~(\ref{th1}). The
expression in (\ref{th6}) is normalized so that for $\Omega=1$ and
$\lambda=0$ we have $D(a)=a$. For arbitrary parameters $(\Omega_0,
\lambda_0)$, expansion of the right-hand side of Eq.~(\ref{th6})
around $a=0$ gives
\begin{equation}    \label{th7}
    D(a) = a + O(a^2).
\end{equation}

\section{Evolution of the overdense region}

We assume that at some time $t_{\rm i}$ corresponding to redshift $z_{\rm
i}$ the region of proper radius $r_{\rm i}$ is overdense by
$\Delta_{\rm i} = {\rm const}$ with respect to the background, that is it
encloses a mass $M(r_{\rm i}) = 4 \pi \rho_{\rm b,i} r_{\rm i}^3
(1+\Delta_{\rm i})/3$, where $\rho_{\rm b,i}$ is the background density at
$t_{\rm i}$.

Evolution of this region is governed by the familiar energy equation
\begin{equation}    \label{th11}
   \frac{1}{2} \left( \frac{{\rm d} r}{{\rm d} t} \right)^2 -
   \frac{G M}{r} - \frac{\Lambda r^2}{6} = E
\end{equation}
which, with a new variable $s=r/r_{\rm i}$, can be
rewritten in the form
\begin{equation}    \label{th12}
    \frac{{\rm d} s}{{\rm d} t} = \frac{H_{\rm i}}{g(s)} , \ \ \
    {\rm where} \ \ \   g(s) = \left[ 1 + \Omega_{\rm i}
    \left( \frac{1}{s} - 1 \right) + \lambda_{\rm i} (s^2 -1)
    \right]^{-1/2}
\end{equation}
and we used the notation $H_{\rm i} = H(z_{\rm i})$, $\Omega_{\rm i} =
\Omega(z_{\rm i})$ and $\lambda_{\rm i} = \lambda(z_{\rm i})$.

Assuming conservation of energy we find that the maximum expansion radius
$r_{\rm ta}$ (or equivalently, $s_{\rm ta} = r_{\rm ta}/r_{\rm i}$)
of the overdense region must obey the condition
\begin{equation}    \label{th14}
    b_1 s_{\rm ta}^3 + b_2 s_{\rm ta} + b_3 = 0
\end{equation}
where $b_1 = \lambda_{\rm i}$, $b_2 = 1 - \Omega_{\rm i} (1+ \Delta_{\rm
i}) - \lambda_{\rm i}$ and $b_3 = \Omega_{\rm i} (1+ \Delta_{\rm i})$.
The only solution to Eq.~(\ref{th14}) which is real, positive and
reproduces the $\lambda_0=0$ case in the limit of small $\lambda_0$ is
\begin{equation}     \label{th15}
     s_{\rm ta} = \frac{2}{\sqrt{3}} \left( \frac{-b_2}{b_1} \right)^{1/2}
     \cos \left(\frac{\phi - 2 \pi}{3} \right)
\end{equation}
with $\phi = \arccos [x/(x^2 + y^2)^{1/2}]$, $x = - 9 b_1^{1/2}
b_3$ and $y = [3 (- 4 b_2^3 - 27 b_1 b_3^2)]^{1/2}$, while for
$\lambda_0=0$ we simply get $s_{\rm ta} = -b_3/b_2$.

The condition for the general solution (\ref{th15}) to be real is
\begin{equation}    \label{th17}
    \Delta_{\rm i} > \Delta_{\rm i, cr} = \frac{1}{\Omega_{\rm i}}
    p(\lambda_{\rm i}) - 1
\end{equation}
where
\begin{equation}    \label{th18}
    p(\lambda_{\rm i}) = 1 + \frac{5 \lambda_{\rm i}}{4}  +
    \frac{3 \lambda_{\rm i} (8 + \lambda_{\rm i})}{4
    q(\lambda_{\rm i})} + \frac{3 q(\lambda_{\rm i})}{4}
\end{equation}
and
\begin{equation}    \label{th19}
    q(\lambda_{\rm i}) = \{ \lambda_{\rm i} [8 - \lambda_{\rm i}^2 + 20
    \lambda_{\rm i} + 8 (1-\lambda_{\rm i})^{3/2}] \}^{1/3}.
\end{equation}
$\Delta_{\rm i, cr}$ is the critical density for the overdense region to
turn around. In the limit of $\lambda_0 \rightarrow 0$ we have
$p(\lambda_{\rm i}) \rightarrow 1$ and we reproduce the well known
condition
\begin{equation}    \label{th20}
    \Delta_{\rm i} > \Delta_{\rm i, cr} (\lambda_0=0) =
    \frac{1}{\Omega_{\rm i}} - 1.
\end{equation}
After turn-around the proper size of the region $r(t)$ evolves almost
independently of the presence of the cosmological constant and can be well
approximated by the analytical solutions for
$\lambda_0=0$: $r/r_{\rm ta} = (1-\cos \theta)/2$ and $t/t_{\rm coll} =
(\theta - \sin \theta)/(2 \pi)$ with $0 \le \theta \le 2 \pi$.

\section{The characteristic densities}

Integrating equations (\ref{th1}) and (\ref{th12}) we get
\begin{equation}
    \int_0^a f(a) {\rm d} a = H_0 t, \ \ \ \
    \int_0^s g(s) {\rm d} s = H_{\rm i} t.   \label{th20a}
\end{equation}
Eliminating $t$ we
obtain equations which can be used to calculate the scale factor at
turn-around ($a_{\rm ta}$) and collapse ($a_{\rm coll}$) of a region with
particular $\Delta_{\rm i}$ at a given $z_{\rm i}$
\begin{eqnarray}
     \int_0^{a_{\rm ta}} f(a) {\rm d} a & = & \frac{H_0}{H_{\rm i}}
     \int_0^{s_{\rm ta}} g(s) {\rm d} s       \label{th21} \\
     \int_0^{a_{\rm coll}} f(a) {\rm d} a & = & 2 \frac{H_0}{H_{\rm i}}
     \int_0^{s_{\rm ta}} g(s) {\rm d} s       \label{th22}
\end{eqnarray}where $s_{\rm ta}$ is given by Eq.~(\ref{th15}).

Assuming that the mass inside the overdense region does not change, the
overdensity inside the sphere of size $r$ with respect to the background
density at any time is
\begin{equation}    \label{th23}
    \delta = \frac{\rho}{\rho_{\rm b}} - 1 = \frac{1}{s^3} \left(
    \frac{a}{a_{\rm i}} \right)^3 (1+ \Delta_{\rm i}) - 1.
\end{equation}
At early times, $t \rightarrow 0$, we can expand the expressions on the
left-hand sides of both equations in (\ref{th20a}) around $a=0$ and
$s=0$ respectively. Integrating term by term we can express time as
a power series of $a$ and $s$ respectively. Inverting those series we
obtain $a$ and $s$ as power series of $t^{2/3}$. Inserting them into
Eq.~(\ref{th23}) and replacing $t$-dependence with $a$ we find
\begin{equation}    \label{th29}
    \delta =  h(\Omega_0, \lambda_0, \Delta_{\rm i}, z_{\rm i}) a + O(a^2),
\end{equation}
where
\begin{equation}   \label{th29a}
    h(\Omega_0, \lambda_0, \Delta_{\rm i}, z_{\rm i}) =
    \frac{3}{5}
    \left[ \frac{1-\Omega_0 -\lambda_0}{\Omega_0} +
    \frac{[\Omega_{\rm i} (1+\Delta_{\rm i}) + \lambda_{\rm i}
    -1](1+z_{\rm i})}{\Omega_{\rm i} (1+\Delta_{\rm i})^{2/3}} \right] .
\end{equation}
Given the behavior of the linear growth factor $D(a)$ in Eq.~(\ref{th7}),
we obtain the density contrast as predicted by linear theory
\begin{equation}    \label{th30}
    \delta_{\rm L} = h(\Omega_0, \lambda_0, \Delta_{\rm i},
    z_{\rm i}) D(a).
\end{equation}

\begin{figure}
\begin{center}
    \leavevmode
    \epsfxsize=11cm
    \epsfbox[50 50 590 310]{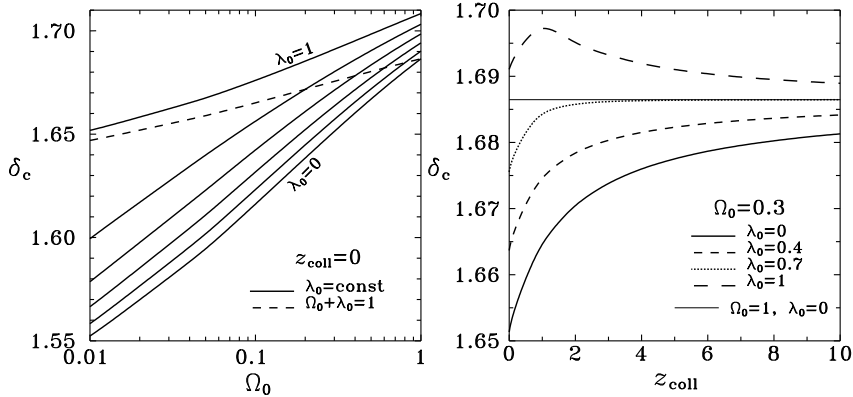}
\end{center}
    \caption{Left panel: parameter $\delta_{\rm c}$ as a function of
    $\Omega_0$ for $z_{\rm coll}=0$. Solid lines correspond from bottom to
    top to $\lambda_0=0, 0.2, 0.4, 0.6, 0.8$ and $1$. The dashed line
    shows results for the flat case $\Omega_0 + \lambda_0 = 1$. Right
    panel: parameter $\delta_{\rm c}$ as a function of $z_{\rm coll}$ for
    four models with $\Omega_0=0.3$ and $\lambda_0=0, 0.4, 0.7$ and $1$.
    The thin straight line marks the fiducial value $\delta_{\rm c} =
    1.68647$ for $\Omega_0 = 1$, $\lambda_0 = 0$.}
\label{d1proc}
\end{figure}

A particularly useful quantity is the linear density contrast at the
moment of collapse i.e. when $s$ reaches zero
\begin{equation}    \label{th31}
    \delta_{\rm c} = h[\Omega_0, \lambda_0, \Delta_{\rm i}(a_{\rm coll}),
    z_{\rm i}] D(a_{\rm coll}).
\end{equation}
$\Delta_{\rm i}(a_{\rm coll})$ in the above equation means that
$\Delta_{\rm i}$ corresponding to $a_{\rm coll}$ has to be determined
numerically for a given $z_{\rm i}$ from the Eq.~(\ref{th22}).
Numerical results for $\delta_{\rm c}$ are shown in Figure~\ref{d1proc}.
In the special case of $\Omega_0 + \lambda_0 = 1$ we reproduce the
results of Eke, Cole \& Frenk~\cite{ecf}, while for open universes with no
cosmological constant our results match those of Lacey \& Cole~\cite{lc}.
Although it is not obvious from Figure~\ref{d1proc}, in the limit of
$\Omega_0 \rightarrow 0$ we have $\delta_{\rm c} \rightarrow 3/2$,
independently of $\lambda_0$.

\begin{figure}
\begin{center}
    \leavevmode
    \epsfxsize=11cm
    \epsfbox[50 50 590 310]{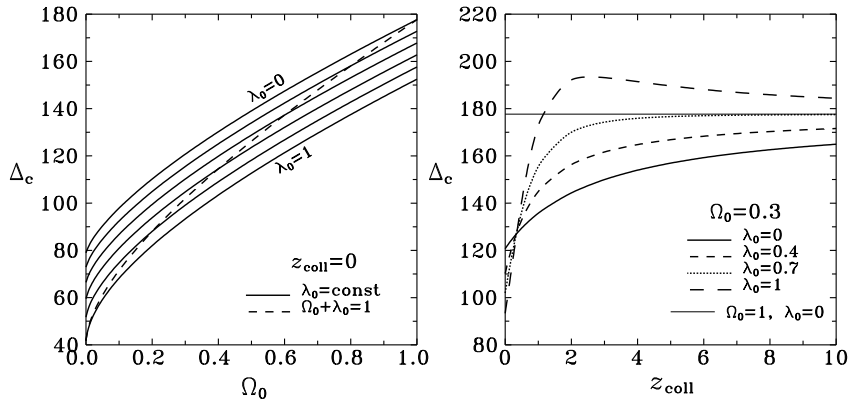}
\end{center}
    \caption{Left panel: parameter $\Delta_{\rm c}$ as a function of
    $\Omega_0$ for $z_{\rm coll}=0$. Solid lines correspond from top to
    bottom to $\lambda_0=0, 0.2, 0.4, 0.6, 0.8$ and $1$. The dashed line
    shows results for the flat case $\Omega_0 + \lambda_0 = 1$. Right
    panel: parameter $\Delta_{\rm c}$ as a function of $z_{\rm coll}$ for
    four models with $\Omega_0=0.3$ and $\lambda_0=0, 0.4, 0.7$ and $1$.
    The thin straight line marks the fiducial value $\Delta_{\rm c} =
    177.653$ for $\Omega_0 = 1$, $\lambda_0 = 0$.} \label{d2proc}
\end{figure}

Another useful quantity is the ratio of the density in the object to the
critical density at virialization
\begin{equation}    \label{th32}
    \Delta_{\rm c} = \frac{\rho_{\rm vir}}{\rho_{\rm crit}} (a_{\rm coll})
    =  \frac{\Omega (a_{\rm coll})}{s_{\rm coll}^3} \left( \frac{a_{\rm
    coll}}{a_{\rm i}} \right)^3 [1+\Delta_{\rm i} (a_{\rm coll})]
\end{equation}
where $s_{\rm coll} = r_{\rm coll}/r_{\rm i}$ and $r_{\rm coll}$ is the
effective final radius of the collapsed object. We assume that the object
virializes at $t_{\rm coll}$, the time corresponding to $s \rightarrow 0$.
Application of the virial theorem in the presence of cosmological constant
leads to the equation for the ratio of the final radius of the
object to its turn-around radius $F=r_{\rm coll}/r_{\rm ta}$ obtained
by Lahav {\it et al}~\cite{llpr}: $2 \eta F^3 - (2 + \eta) F + 1 = 0$
where $\eta = 2 \lambda_{\rm i} s_{\rm ta}^3/[\Omega_{\rm i}(1+\Delta_{\rm
i})]$. Numerical results for $\Delta_{\rm c}$ in different models are
shown in Figure~\ref{d2proc}. Again, we agree with the results for the
special cases of $\Omega_0<1$, $\lambda_0=0$ and $\Omega_0 + \lambda_0=1$
derived previously by Lacey \& Cole~\cite{lc} and Eke {\it et
al}~\cite{ecf} respectively. It should be emphasized that the value of
$\Delta_{\rm c}$ is strongly model-dependent, contrary to the common
assumption of $\Delta_{\rm c} \approx 200$.

\section{The characteristic redshifts}

\begin{figure}
\begin{center}
    \leavevmode
    \epsfxsize=11cm
    \epsfbox[50 50 590 310]{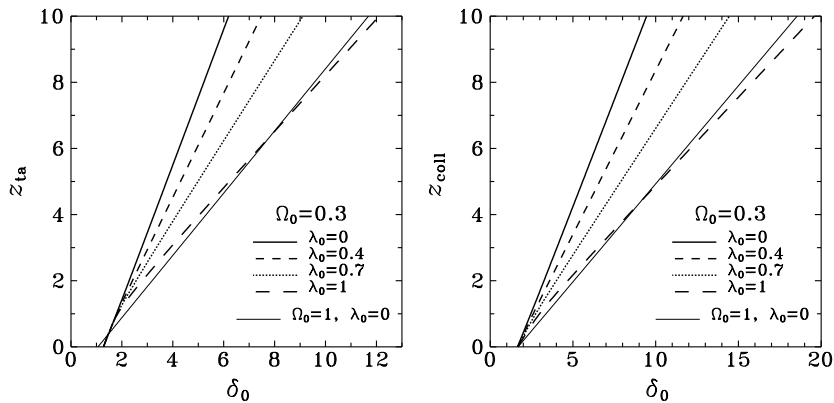}
\end{center}
    \caption{Redshift of turn-around (left panel) and collapse (right
    panel) of density fluctuation with present density contrast $\delta_0$
    for four models with $\Omega_0=0.3$ and $\lambda_0=0, 0.4, 0.7$ and
    $1$. The thin straight line marks the fiducial case of $\Omega_0 = 1$,
    $\lambda_0 = 0$.}
\label{zall}
\end{figure}

It is sometimes useful to be able to estimate the redshift of a particular
stage of evolution of perturbation given its present overdensity as
predicted by linear theory, $\delta_{\rm L} (a=1) = \delta_0$.
For the redshift of collapse combining equations (\ref{th30}) and
(\ref{th31}) gives $\delta_0 = \delta_{\rm c} (a_{\rm coll})
D(a=1)/D(a_{\rm coll})$.
Using the previously obtained results for $\delta_{\rm c}$ and
formula (\ref{th6}), we can calculate the present linear density contrast of
fluctuation that collapsed at $z_{\rm coll}$. This relation can only be
inverted analytically in the case of $\Omega_0=1$, $\lambda_0=0$ when we
get $z_{\rm coll} = \delta_0/\delta_{\rm c} -1$. In an analogous way one
can obtain the redshifts of turn-around, $z_{\rm ta}$. Numerical
results for both quantities are shown in Figure~\ref{zall}. Useful fitting
formulae for those results will be presented in our forthcoming
paper~\cite{lh}.

%\vspace{-0.1in}

\section*{Acknowledgments}

This work was supported by the Polish KBN grant 2P03D02319 and by
the Israel Science Foundation grant 103/98.

%\vspace{-0.1in}

\end{document}